\documentclass{icrc2009}

\usepackage{graphicx}   % for including figures
\usepackage[caption=false]{caption}    % for captions
\usepackage[font=footnotesize]{subfig} % subfig.sty for a double column floating figure using two subfigures
\usepackage{fixltx2e}
\usepackage{url}

\newcommand{\shorttitle}[1]%
{\markboth{Proceedings of the 31\MakeLowercase{$^{st}$} ICRC, {\L}\'{o}d\'{z} 2009}{#1} }
\newcommand{\etal}{\MakeLowercase{\textit{et al. }}} % "et al."

\begin{document}
\title{Search for neutrino flares from point sources with IceCube}

\author{\IEEEauthorblockN{J. L. Bazo Alba\IEEEauthorrefmark{1}, E. Bernardini\IEEEauthorrefmark{1}, R. Lauer\IEEEauthorrefmark{1}, for the IceCube Collaboration\IEEEauthorrefmark{2} }
                            \\
\IEEEauthorblockA{\IEEEauthorrefmark{1}DESY, D-15738 Zeuthen, Germany.}
\IEEEauthorblockA{\IEEEauthorrefmark{2} see 
http://www.icecube.wisc.edu/collaboration/authorlists/2009/4.html}
}

% please write the preseter's name and short title (3-4 words maximum)
%    which will appear at the header of the even pages.
\shorttitle{J. L. Bazo Alba \etal Neutrino Flares with IceCube}
\maketitle

\begin{abstract}
 A time-dependent search for neutrino flares from pre-defined directions in the whole sky is presented. The analysis uses a time clustering algorithm combined with an unbinned likelihood method. This algorithm provides a search for significant neutrino flares over time-scales that are not fixed a-priori and that are not triggered by multiwavelength observations. The event selection is optimized to maximize the discovery potential, taking into account different time-scales of source activity and background rates. The method is applied to a pre-defined list of bright and variable astrophysical sources using 22-string IceCube data. No significant excess is found.
\end{abstract}

 
\section{Introduction}
 Several astrophysical sources are known to have a variable photon flux at different wavelengths, showing flares that last between several minutes to several days. Hadronic models of Active Galactic Nuclei (AGNs) predict \cite{Torres}\cite{Dermer} neutrino emission associated with these multiwavelength (MWL) emissions. Time integrated analyses are less sensitive in this flaring scenario because they contain a higher background of atmospheric neutrinos and atmospheric muons. Therefore a time dependent analysis is more sensitive because it reduces the background by searching smaller time scales around the flare. A direct approach that looks for this correlation using specific MWL observations is reported in \cite{Baker}. 

In order to make the flare search more general, and since MWL observations are scarce and not available for all sources, we take an approach not triggered by MWL observations. We apply a time-clustering algorithm (see \cite{Satalecka}) to pre-defined source directions looking for the most significant accumulation in time (flare) of neutrino events over  background, considering all possible combinations of event times. One disadvantage of this analysis is the increased number of trials, which reduces the significance. Nevertheless, for flares sufficiently shorter than the total observation period, the time clustering algorithm is more sensitive than a time integrated analysis. The predicted time scales are well below this threshold.\\

\section{Flare search algorithm}
The time clustering algorithm chooses the most promising flare time windows based on the times of the most signal-like events from the analyzed data. Each combination of these event times defines a search time window ($\Delta t_i$). For each $\Delta t_i$ a significance parameter $\lambda_i$ is calculated. The algorithm returns the best $\lambda_{max}$ corresponding to the most significant cluster. The significance can be obtained using two approaches: a binned method, as in the previous implementation \cite{Satalecka}, and an improved unbinned maximum likelihood method \cite{Braun} which enhances the performance.

The unbinned maximum likelihood method defines the significance parameter by: 

\begin{equation}
\lambda=-2 \log\left[ \frac{\mathcal{L}(\vec{x}_{s},n_s=0)}{\mathcal{L}(\vec{x}_{s},\hat{n}_s,\hat{\gamma}_s)}  \right],
\label{lambda}
\end{equation}
where $\vec{x}_{s}$ is the source location, $\hat{n}_s$ and $\hat{\gamma}_s$ are the best estimates of the number of signal events and source spectral index, respectively, which are found by maximizing the likelihood, ($\mathcal{L}$):

\begin{equation}
 {\cal L}=\prod_{i=1}^{n_{tot}} \left ( \frac{n_{s}}{n_{tot}}S_{i}+ \left( 1-\frac{n_{s}}{n_{tot}} \right) B_i \right )
 \label{llh_unbinned}
\end{equation}

\begin{figure}[h]
\centering
\includegraphics[width=2.5in]{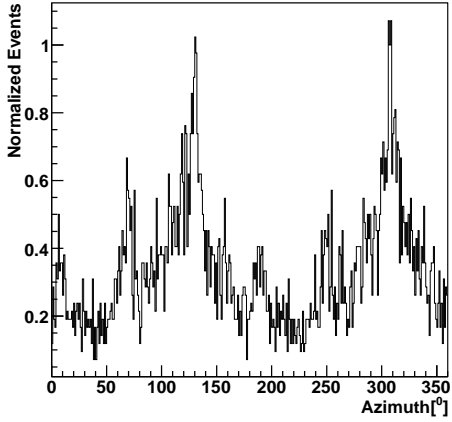}
\caption{Normalized azimuth distribution of the data sample reported in \cite{Bazo}.}
\label{plot_azimuth}
\end{figure}

The background probability density function (pdf), $B_i$, calculated purely from data distributions, is given by: 
\begin{equation}
B_{i}=P^{space}_{i}(\theta_{i},\phi_{i})P^{energy}_{i}(E_{i},\theta_{i})P^{time}_{i}(t_{i},\theta_{i}),
 \label{llh_bg}
\end{equation}
where $P^{space}_{i}$ describes the distribution of events in a given area (a zenith band of 8$^{\circ}$ is used for convenience). In a simple case this probability would be flat because of random distribution of background events. However, due to applied cuts, Earth absorption properties and detector geometry, this probability is dependent on zenith, $\theta_i$, and azimuth, $\phi_i$. The irregular azimuthal distribution caused by the detector geometry is shown in Fig.~\ref{plot_azimuth}. For time integrated analyses covering one year the dependence on the azimuth is negligible because the exposure for all right ascension directions is integrated.  However, an azimuth correction becomes important for time scales shorter than 1 day, reaching up to 40\% difference, thus it should be included in time dependent analyses. $P^{space}$ has value unity when integrated over solid angle inside the test region (i.e. zenith band). 

\begin{figure}[h]
\centering
\includegraphics[width=3.1in]{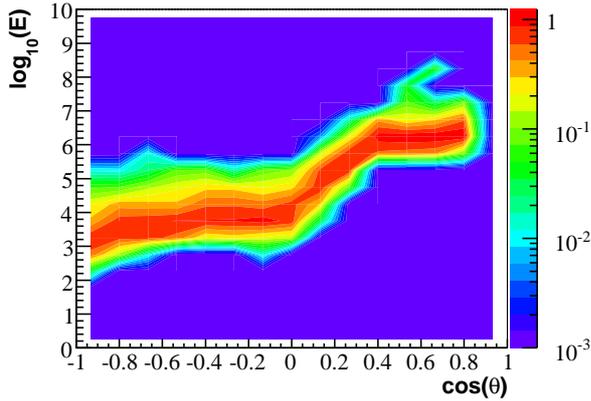}
\caption{Background energy pdf from data as a function of the energy estimator and zenith angle. The ultra high energy sample \cite{Lauer} is used. The southern sky corresponds to $\cos(\theta)>0$.}
\label{plot_CosMue_UHE}
\end{figure}

The energy probability $P^{energy}_i$  is determined from the energy estimator distribution and depends on the zenith coordinate. In the southern sky an energy sensitive event selection is the most efficient way to reduce the atmospheric muon background. This energy cut decreases with zenith angle, thus creating a zenith dependence of the energy. Therefore a zenith dependent energy probability, shown in Fig.~\ref{plot_CosMue_UHE}, is needed. Note that for the northern sky this correction is small. 

\begin{figure}[h]
\centering
\includegraphics[width=3.3in]{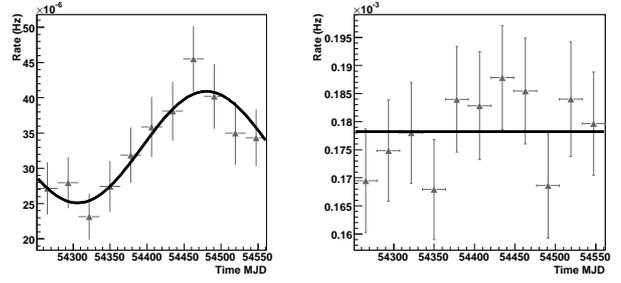}
\caption{Uptime corrected rates and their fits for the southern (left) and northern (right) skies.}
\label{plot_fitrate}
\end{figure}

The time probability $P^{time}_i$ is described by fits to the event rates in the entire observed period as a function of time ($t_{i}$). Two regions of the sky (South and North) are distinguished because they have different properties, thus the zenith, $\theta_i$, dependence of $P^{time}_i$. The northern sky sample consists mostly of atmospheric neutrinos which do not show a significant seasonal variation, therefore a constant fit is used. For the southern sky, a sinusoidal fit is used because it is dominated by a background of high energy atmospheric muons which have seasonal variation. These fits are shown in Fig.~\ref{plot_fitrate} and include the necessary correction for the uptime\footnote{The uptime takes into account the inefficiency periods and data gaps after data quality selection.} of the detector. It has been verified that the time modulations for different zenith bands within a half hemisphere are the same, thus allowing us to use all events inside the half hemisphere for the fit of the rates. 

\begin{figure}[h]
\centering
\includegraphics[width=3.in]{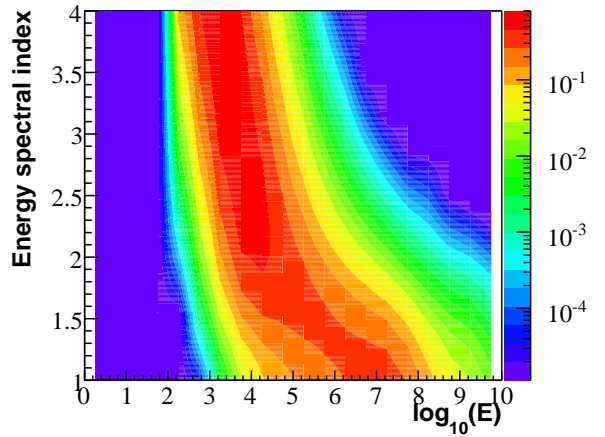}
\caption{Projection for the whole sky of the energy component of the signal pdf as a function of the energy estimator and energy spectral index. The ultra high energy sample \cite{Lauer} is used.}
\label{plot_E_gamma_UHE}
\end{figure}

The signal pdf, $S_{i}$, is given by: 
\begin{equation}
S_{i}=P^{space}_i(\mid \vec{x}_i-\vec{x}_{s} \mid, \sigma_{i})P^{energy}_i(E_{i},\theta_{i},\gamma_{s})P^{time}_{i},
\label{llh_signal}
\end{equation}
where, the spatial probability, $P^{space}_i$ is a Gaussian function of $\mid \vec{x}_i-\vec{x}_{s} \mid$, the space angular difference between the source location, $\vec{x}_{s}$, and each event's reconstructed direction, $\vec{x}_{i}$, and $\sigma_{i}$, the angular error estimation of the reconstructed track. The estimator used for $\sigma_{i}$ is the size of the error ellipse around the maximum value of the reconstructed event track likelihood.  The energy probability, $P^{energy}_i$, constructed from signal simulation, is a function of the event energy estimation, $E_i$, the zenith coordinate, $\theta_i$, and the assumed energy spectral index of the source, $\gamma_s$ ($E^{-\gamma_s}$). A projection of $P^{energy}_i$ for the whole sky is shown in Fig.~\ref{plot_E_gamma_UHE}. For a given $\theta_i$ and $\gamma_s$ the energy pdf is normalized to unity over $E_i$. For the energy a dedicated estimator of the number of photons per track length is used. The time probability, $P^{time}_{i}$, is constant since no flare time structure is assumed (i.e. taken to be flat in time).

\begin{figure*}[!t]
  \centerline{\subfloat[Southern sky at (dec=-7.6, ra=306.4)]{\includegraphics[width=2.5in]{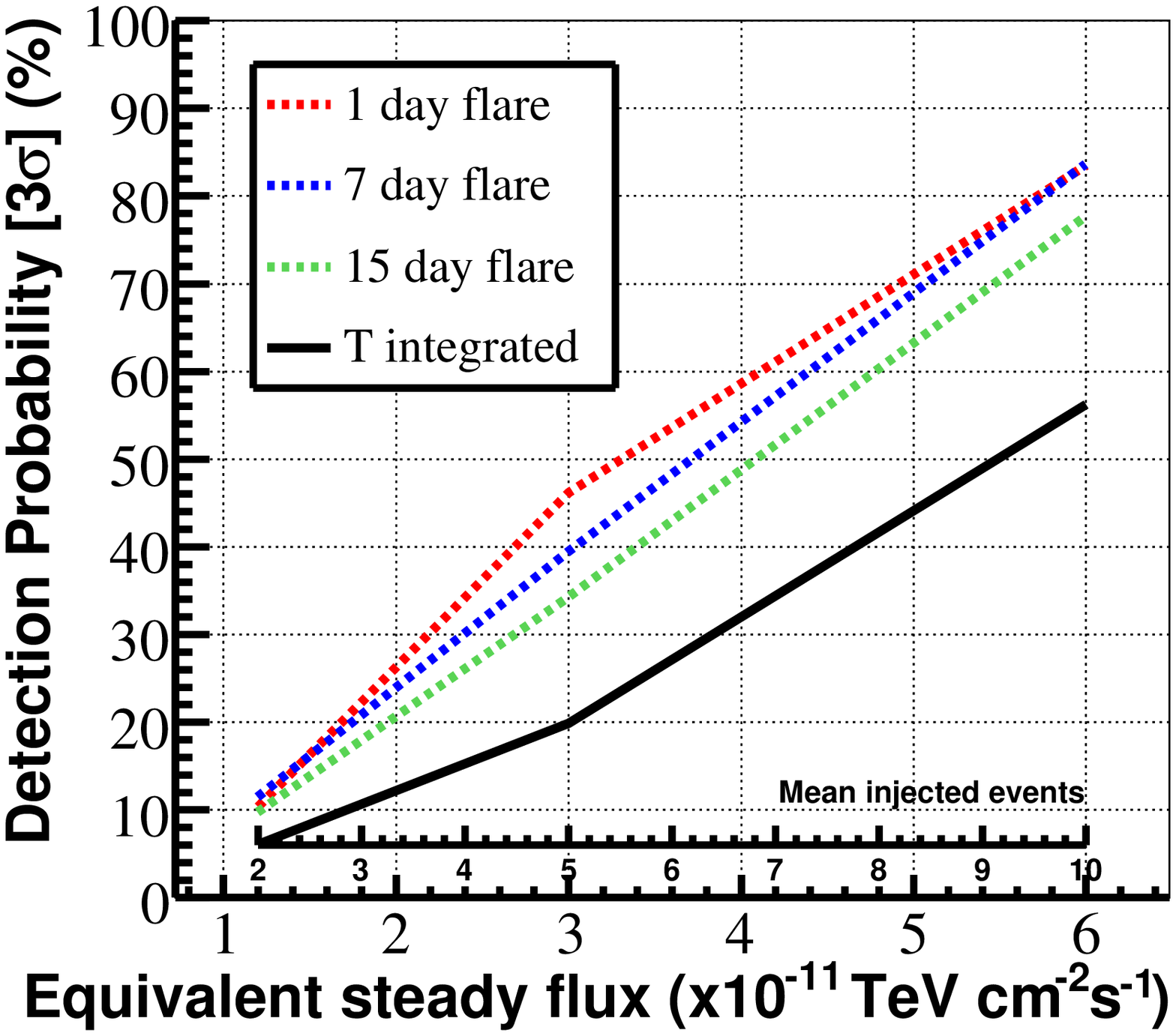} \label{sub_fig1}}
    \hfil
    \subfloat[Northern sky at (dec=16.1, ra=343.5)]{\includegraphics[width=2.5in]{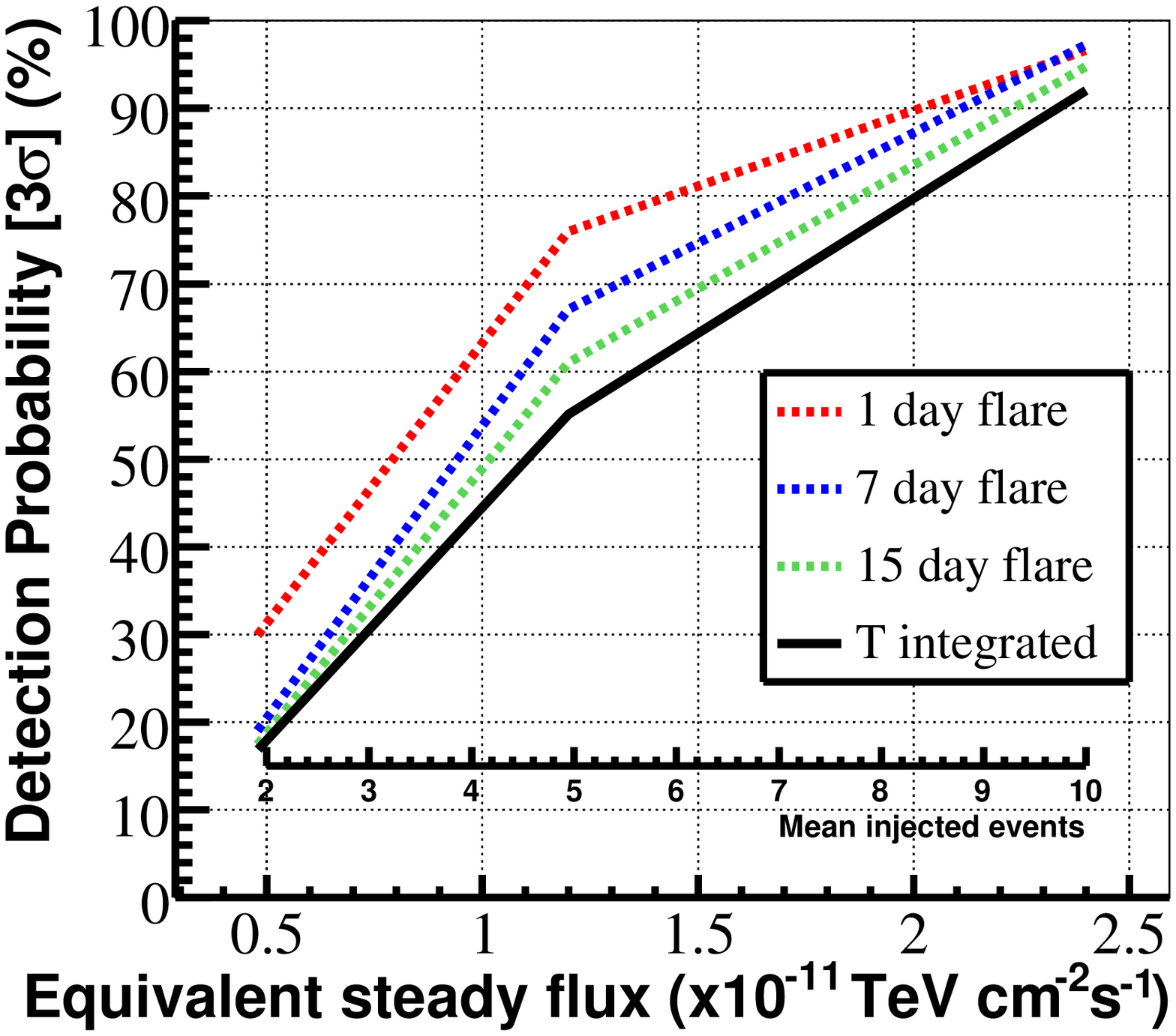} \label{sub_fig2}}
  }
  \caption{Detection probability (3$\sigma$) for two source directions. The curves correspond to different time duration of the flares as function of the injected flux with a $E^{-2}$ energy spectrum, using an unbinned time variable method (dashed), compared to a time integrated method (solid). The same mean number of events are injected into the time-windows (1, 7, 15, and 276 days) at each point on the x-axis, which is labeled with the equivalent flux corresponding to the full 276 day period.} 
  
  \label{plot_detprob}
\end{figure*}

We use a binned method implementation of the time clustering algorithm as a crosscheck of our new unbinned analysis. In the case of the binned method, a circular angular search bin ($2.5^\circ$ radius) around the source direction is used. The times of the events that define the search time windows ($\Delta t_i$) are given by all the events inside this angular bin. The significance parameter is obtained from Poisson statistics, given the number of expected background events inside the bin and the observed events in each cluster with multiplicity\footnote{The integral of the Poisson distribution of the background events starts at (m-1) since the beginning and end of the time period are fixed from the data itself.} m. The expected number of background events is calculated by integrating, in the given time window, the fit to the rates, as described above. This calculation takes into account the zenith dependence of the background, in zenith bands with the size of the bin, the corresponding uptime factor and the azimuth correction.

The best significance obtained for a cluster is corrected for trial factors by running several Monte Carlo background-only simulations. The simulation is done by creating distributions from data of zenith, azimuth, reconstruction error and energy estimator. The event characteristics are randomly taken from these distributions while considering the correlations between the different parameters. In order to study the performance of the algorithm, we calculate the neutrino flare detection probability as a function of the signal strength and duration of the flare by simulating signal events on top of background events\footnote{The number of injected background and signal events is Poisson distributed.}. The properties of signal events are taken from a dedicated signal simulation and depend on the assumed energy spectral index. The Point Spread Function (PSF) is used to smear the events around the source location, thus simulating the effect of the direction reconstruction. For each simulation, a random time is chosen around which signal events are randomly injected inside the time window defined by the flare duration. The flare duration is investigated in the range from 1 day to 15 days, though the algorithm finds the best time window, which could be larger. We constrain the largest flare duration in the algorithm to be less than 30 days, which is sensible from $\gamma$-ray observations. 

\section{Source selection}
Since searching for all directions in the sky would decrease the significance, we consider only a few promising sources, thus reducing the number of trials. We select variable bright astrophysical sources in the whole sky. The selected blazars, including Flat Spectrum Radio Quasars (FSRQs) and Low-frequency peaked BL Lacs (LBLs), are taken from the confirmed Active Galactic Nuclei (AGN) in the third EGRET catalogue (3EG) \cite{EGRET}. We also require that they are present in the current latest Fermi catalogue (0FGL) \cite{Fermi}. The criteria for selecting variable and bright source is based on the following parameters thresholds:

\begin{itemize}
\item Variability index (3EG) $> 1$
\item Maximum 3EG flux ($E> 100$ MeV) $> 40$ [$10^{-8}$ ph cm$^{-2}$s$^{-1}$]
\item Average 3EG flux ($E> 100$ MeV) $> 15$ [$10^{-8}$ ph cm$^{-2}$s$^{-1}$]
\item Inside visibility region of IceCube.
\end{itemize}

The selected source list consists of 10 directions (Table~\ref{table_results}) that are going to be tested with the time clustering algorithm. Models like \cite{Dermer} favor fluxes of higher energy neutrinos from FSRQ sources. Given the absorption of neutrinos at different energies in the Earth and the event cut strategy, southern sky FSRQs are more favored by these models because of their higher energy range of sensitivity.

\begin{table*}[th]
\caption{Results for pre-defined variable astrophysical source candidates using the time clustering algorithm. }
\label{table_results}
\centering
\begin{tabular}{|c|c|c|c|c|c|c|}
\hline
Source & Type & dec [$^\circ$] & ra [$^\circ$]&  p-value & $\Delta t$ (days)& Fluence Limit (GeV/cm$^{2}$) \\

\hline
\hline
GEV J0540-4359 & LBL  & -44.1 & 84.7   & 0.54  & 7.08 &  29.8 \\
GEV J1626-2502 & FSRQ & -25.5 & 246.4  & 0.41  & 22.8 &  22.8 \\
GEV J1832-2128 & FSRQ & -21.1 & 278.4  & 0.64  & 4.49 &  12.0 \\
GEV J2024-0812 & FSRQ & -7.6  & 306.4  & 1     & 3.55 &  3.7 \\
3C 279         & FSRQ & -5.8  & 194.1  & 0.52  & 0.19 &  3.3 \\
3C 273         & FSRQ &  2.0  & 187.3  & 0.84  & 1.97 &  0.37 \\
CTA 102        & FSRQ & 11.7   & 338.1 & 0.27  & 3.7  &  1.42 \\
GEV J0530+1340 & FSRQ & 13.5  &  82.7  & 0.82  & 3.48 &  0.47 \\
3C 454.3       & FSRQ & 16.1  & 343.5  & 0.047  & 0.5 &  2.22 \\
GEV J0237+1648 & LBL  & 16.6  &  39.7  & 0.59  & 1.99 &  1.08 \\
\hline
\end{tabular}
\\
 \vspace{0.3cm}
Note.- The flare duration of the best cluster is given by $\Delta t$. The fluence upper limit is calculated by integrating $d\Phi/dE \times E$ over the 90\% energy range and $\Delta t$, assuming a neutrino energy spectrum of $E^{-2}$.

\end{table*}

\section{Data Samples}

IceCube\cite{IceCube} 22-string data from 2007-08 is used. It spans 310 days with an overall effective detector uptime of 88.9\% (i.e. 276 days). The whole sky (declination range from -50$^{\circ}$ to 85$^{\circ}$) is scanned. Different selection criteria are applied for the northern and southern skies. Previously obtained reconstructed datasets are used: the standard point source sample for the northern sky \cite{Bazo} (5114 events, declination from -5$^{\circ}$ to 85$^{\circ}$, $1.4^{\circ}$ sky-averaged median angular resolution) and the dedicated ultra high energy sample for the southern sky \cite{Lauer} (1877 events in the whole sky, declination from -50$^{\circ}$ to 85$^{\circ}$, $1.3^{\circ}$ sky-averaged median angular resolution). The first sample is optimized, within an unbinned method, for the optimal sensitivity to both hard and soft spectrum sources. The second sample was optimized for a binned method at ultra high energies. Therefore it should be noted that the binned method results are much better in the southern sky than in the northern sky. Nevertheless, the unbinned method, for an $E^{-2}$ energy spectrum still performs better in the southern sky.

 The energy containment in these two regions is different, with ranges from TeV to PeV and from PeV to EeV, in the northern and southern sky respectively. Event tracks are obtained with a multi-photoelectron \footnote{The MPE reconstruction takes the arrival time distribution of the first of N photons using the cumulative distribution of the single photon pdf.} (MPE) \cite{AMANDAreco} reconstruction which improves the angular resolution for high energies. 

\section{Results}
The probability of a $3\sigma$ flare detection using this time variable analysis (time clustering algorithm) for a given number of Poisson mean injected signal events with a $E^{-2}$ energy spectrum is shown for two sources, at the southern and northern skies, for different time scales in Fig.~\ref{plot_detprob}. For comparison purposes, time-integrated detection probabilities integrated over the whole 22-string IceCube data period (276 days) are also given. For shorter flare durations the detection probability increases and is well above a time integrated search. In the northern sky, the same simulated signal of 5 mean injected events in a 7-day window was on average 1.3 times more likely to be detected at $3\sigma$ with the unbinned time variable search than with the time integrated search, and in the southern sky, on average about twice as likely with the time variable search. 

The time clustering algorithm was applied to the selected sources candidates. No significant excess above the atmospheric background is found, therefore upper limits on the neutrino fluence are calculated. The results are presented in Table~\ref{table_results}. The highest fluctuation observed corresponds to 3C 454.3 with a p-value of 4.7\% (not including the trial factors due to looking at several sources).

\section{Summary}

We have presented the sensitivity of the time clustering algorithm using an unbinned maximum likelihood method. This is an improvement over the previous performances using a binned method and time integrated analyses. The search window for variable sources has been extended to the southern sky. IceCube 22-string data was analyzed using this method looking for neutrinos flares with no a priori assumption on the time structure of the signal. Since no deviation from the background-only hypothesis was found, upper limits on the neutrino fluence from these sources were derived.

\end{document}